\documentclass{elsart}

\usepackage{harvard}

\usepackage{graphicx}

\usepackage{amssymb}

\def\url#1{{\ttfamily\def\/{/\discretionary{}{}{}}#1}}

\def\ltsima{$\; \buildrel < \over \sim \;$}
\def\simlt{\lower.5ex\hbox{\ltsima}} 
\def\gtsima{$\; \buildrel > \over \sim \;$}
\def\simgt{\lower.5ex\hbox{\gtsima}} 

\begin{document}

\begin{frontmatter}
\title{{\it Hubble Space Telescope} Observations of BL Lacertae
Environments\thanksref{label1}}

\thanks[label1]{Based on observations made
with the NASA/ESA Hubble Space Telescope, obtained at the Space
Telescope Science Institute, which is operated by the Association of
Universities for Research in Astronomy, Inc., under NASA contract
NAS~5-26555.}

\author[pesce]{Joseph E. Pesce}, 
\author[urry]{C. Megan Urry},
\author[urry]{Matthew O'Dowd},
\author[urry]{Riccardo Scarpa},
\author[falomo]{Renato Falomo},
\author[treves]{Aldo Treves}

\address[pesce]{Eureka Scientific}
\address[urry]{Space Telescope Science Institute}
\address[falomo]{Osservatorio Astronomico di Padova}
\address[treves]{University of Milan at Como}

\begin{abstract}

\noindent We analyze images of BL Lacertae objects obtained with the
{\it Hubble Space Telescope} WFPC2 and the F814W filter.  The nine
objects cover a redshift range of 0.19 to 0.997.  The relatively
deep images are sufficient to detect galaxies at least one magnitude
below M$^{*}_{I}$ (--21.4) and in most cases to three magnitudes below M$^{*}$.
Galaxy enhancement over the average background is found around four out
of the nine objects.  Results for some cases are confirmed by ground-based 
imaging. In the other cases, the redshifts of the target
BL Lac objects may be incorrect or they are truly isolated.  These findings
reinforce the idea that on average, BL Lac objects are found in regions
of above average galaxy density.  However, isolated objects apparently can host
BL Lac nuclei too, a result that has implications for the processes
that trigger/fuel the nuclear activity.

\end{abstract}

\end{frontmatter}

\section{Introduction}
\label{intro}

BL Lacertae objects (BL Lacs) are an extreme form of active galactic
nuclei (AGN) which exhibit rapid flux variability at all frequency,
high polarization, and weak or non-existent spectral features.  Along
with their cousins, the Flat-Spectrum Radio Quasars, they form the
blazar class of AGN.  The model which best describes blazars is one in
which a jet of relativistic material is beamed directly at us
\cite{bla78}.

Although among the rarest types of AGN, BL Lacs are particularly
important precisely because we can look into their jets.  In this way,
we can see the site of energy production which is presumably close to
the central engine, or black hole.  Thus, by studying these enigmatic
objects we can better understand the central black hole in all other
AGN.

One way of studying these objects, which is independent of any
assumptions made about the nuclear regions and energy production, is
by their environments.  There is growing consensus that BL Lacs are
found on average in poor clusters of galaxies \cite{fal93,fal95,sti93,pes94,pes95,smi95,wur97}. However, some individual objects (e.g. PKS 0548--322) are found in
rich clusters \cite{fal95}.

Work has progressed on the kilo-parsec-scale environments, or host
galaxies of BL Lacs as well.  BL Lacs are found in giant elliptical
galaxies with $-21.5 \simlt M_V \simlt -24.5$~mag \cite{abr91,fal96,wur96,fal97,urr99}. Some individual objects were thought to
be hosted by spiral galaxies, but high resolution observations have
shown them to be ellipticals \cite{urr99}.

High resolution images of the host galaxies of six radio-selected BL
Lac objects observed with the {\it Hubble Space Telescope (HST)} have
been presented in \citeasnoun{fal97} and \citeasnoun{urr99}.  These
are relatively deep WFPC2 images from cycle 5.  In addition, we are
currently analyzing the host galaxies and extended environments of
more than 100 BL Lac objects at redshifts 0.03 - 1 observed by HST as
part of a snapshot project (Pesce et al. 2000, in preparation; Scarpa
et al. 2000 in preparation; Urry et al.  2000, in preparation).  The
combined dataset will allow a thorough investigation of BL Lac
environmental properties over a large redshift range.

In this paper we present the analysis of the extended environments of
six radio selected objects from our HST cycle 5 GO project (PI URRY).  To
this sample we have added three X-ray selected objects obtained from
the HST archive and observed during cycle 5 (PI JANNUZI). We assume
H$_0 = 50$ km s$^{-1}$ Mpc$^-1$ and q$_0 = 0.5$.

\section{Data Reduction \& Analysis}
\label{analysis}

The objects were observed with the WFPC2 and the F814W (I-band)
filter, the BL Lac was centered in PC camera. Exposure times varied and were approximately 10 min to 1.5 hours (see Table 1).  More details about the
reduction and analysis of these fields can be found in \citeasnoun{urr99}.

We used the FOCAS software, in IRAF, to detect and classify all objects on 
an image and produce a catalog (Figure 1). For the central BL Lac, the 
Point Spread Function (PSF) was created with
Tiny Tim \cite{has95}; each object is classified as `Star' or
`Galaxy' based on an automatic comparison of its profile to a standard
PSF profile. In these images, we detect objects down to a ``completeness 
limit'' of $m = 24$.  Objects fainter than the completeness limit are not 
considered in the analysis of the images. 

The automatic identifications from FOCAS were checked with
simulated images consisting of stars, galaxies, and noise.  FOCAS
detected and correctly classified all objects above the completeness
limit and 90\% of objects below the limit. As a further check, the
galaxy density on each image was compared to the average
background density for the F814W filter from Casertano (1997, priv com). 

\begin{figure}
\begin{center}
\includegraphics*[width=10cm]{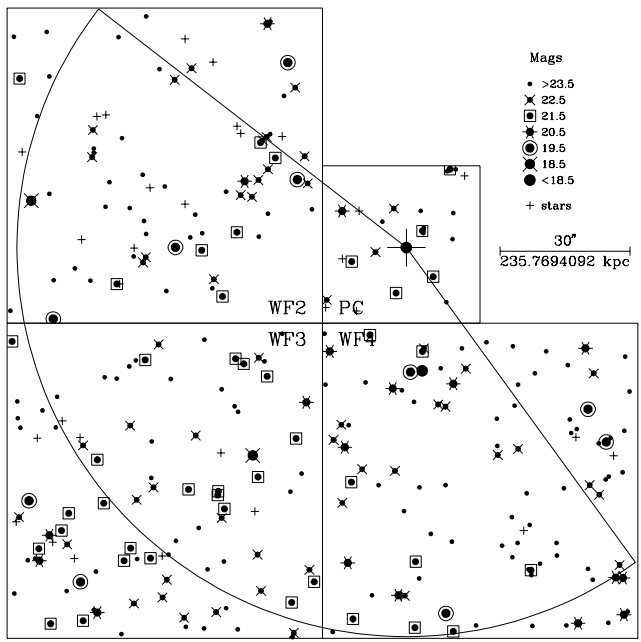}
\end{center}
\caption{A map of the region around 1823+568, as
seen by HST.  The Planetary Camera is labeled as PC while the three
wide field chips are WF2, WF3, and WF4.  The object is marked by the
large cross centered in the PC and the other symbols are explained in
the key.  For this object, $z = 0.664$ and the PC is $\sim$280 kpc
square, while the maximum complete radius sampled (the straight lines
and arc) is $\sim$0.7 Mpc.}
\label{fig_fig1}
\end{figure}


\begin{table*}
\caption{BL Lac Target Objects.}
\label{tab_bll} 
\begin{center}
\begin{tabular}{c c r c}
\hline 
Object & $z$ & Exp (sec) & $m^*_{\rm I}$$^{\#}$ \\
\hline 
0814+425 & 0.258 & ~~475& 19.7 \\
0828+493 & 0.548 & 2060 & 21.4 \\
1221+249 & 0.218 & 1900 & 19.3 \\
1308+326 & 0.997 & 1800 & 22.8 \\
1407+599 & 0.495 & 4080 & 21.2 \\
1538+149 & 0.605 & 1670 & 21.6 \\
1823+568 & 0.664 & 2120 & 21.9 \\
2143+070 & 0.237 & ~~850& 19.5 \\
2254+074 & 0.190 & 1480 & 19.0 \\
\hline 
\end{tabular}
\end{center}
\vspace*{.6cm}
\noindent
$^{\#}$ Apparent magnitude of a typical galaxy, assuming $M^*_{\rm I} = -21.4$.  No K-correction has been applied. \\ 
\end{table*}

\section{Results}
\label{results}

To avoid background contamination, we only count
galaxies with $m_I \leq m_I^* +1$ where $M_I^* = -21.4$. 
The PC images subtend 0.3 arcmin$^2$ while the WF chips
subtend 1.5 arcmin$^2$ giving a maximum radial extent from our BL Lacs
of 0.4 - 0.8 Mpc.  Most BL Lac clusters are found within 0.5 Mpc of
object \cite{pes94,pes95}. However, because of the WFPC2 geometry, 
we only observe $\sim$1/2 of the volume surrounding the BL Lac.

We checked our procedures by analyzing an archival image of
the rich cluster Abell 2390 (richness class 1, $z = 0.231$, exposure =
2100 s).  On this image, we detected a significant (factor 9-20) overdensity
of galaxies.

For our objects, we {\bf detect enhancements (factor 2-4) above
background in four cases:} 1407+599, 1538+149, 1823+568, 2143+070.
A modest enhancement is found around 2254+074, but this object has the
lowest redshift so we are observing only a small part of the potential
cluster volume.

{\bf No enhancements} are found for 0814+425, 0828+493,
1221+249, 1308+326. Given our completeness limit, and assuming correct 
redshifts, we {\it should} detect any
cluster present (if the cluster is approximately symmetric). {\bf These
objects appear to be truly isolated.}  

\section{Conclusions}
\label{conclusions}

We have analyzed the extended environments of nine BL Lac objects
observed with the HST WFPC2 in the F814W filter.  The long exposures
provide deep ($m \sim 24$ mag) images.  We find significant
enhancement of galaxy density around four of the nine BL Lac objects
(1407+599, 1538+149, 1823+568, 2143+070).  In some cases (1407+599,
1823+568) the enhancement is extreme, with dozens of galaxies within
$\sim$150 kpc of the BL Lac object.

The remaining five fields show no indication of	excess galaxy counts.
In fact, some fields are slightly below the average background galaxy
density, although in general the galaxy density is consistent with the
background.  These objects are either truly isolated, their clusters
are unusually asymmetric (and strangely all out of the field of view),
or the redshifts are incorrect.  Of these three explanations, the
first is most likely.  Thus, some BL Lac objects appear to be in
regions of significantly enhanced galaxy density, while others seem
to be completely isolated.  This result is not surprising given that
BL Lac hosts appear to be otherwise normal giant elliptical galaxies.
Such galaxies are found in all types of environments - from rich
clusters to isolated regions of space.  Nonetheless if galaxy
interactions play an important role in the AGN activity, future work
needs to address this point.  Our sample is too small to determine
differences between X-ray and radio selected objects, a point to be
addressed in our analysis of the larger HST snapshot survey.


\end{document}